%% file: discreteNN.tex
\documentclass[twoside,aps,10pt,twocolumn]{revtex4-1}

\usepackage{amsmath, amssymb}
\usepackage{bm}
\usepackage[T1]{fontenc} 	
\usepackage[utf8]{inputenc}
\usepackage{graphicx, float}	
\usepackage{subfigure}		
\usepackage{geometry}		
\usepackage{url}
\usepackage{braket}
\usepackage{cancel}
\usepackage{color}
\usepackage[breaklinks=true,colorlinks=true,allcolors=blue]{hyperref}

\begin{document}
\title{Reconstruction of nanoscale particles from 
single-shot wide-angle FEL diffraction patterns with physics-informed neural 
networks}
\author{Thomas Stielow}
\author{Stefan Scheel}
\affiliation{Institut f\"ur Physik, Universit\"at Rostock,
Albert-Einstein-Stra{\ss}e 23--24, D-18059 Rostock, Germany}
\date{\today}

\begin{abstract}
Single-shot wide-angle diffraction imaging is a widely used method to 
investigate the structure of non-crystallizing objects such as nanoclusters, 
large proteins or even viruses. Its main advantage is that information about 
the three-dimensional structure of the object is already contained in a single 
image. This makes it useful for the reconstruction of fragile and 
non-reproducible particles without the need for tomographic measurements. 
However, currently there is no efficient numerical inversion algorithm 
available that is capable of determining the object's structure in real time. 
Neural networks, on the other hand, excel in image processing tasks suited for 
such purpose. Here we show how a physics-informed deep neural network can be 
used to reconstruct complete three-dimensional object models 
of uniform, convex particles on a voxel grid 
from single two-dimensional wide-angle scattering patterns. We demonstrate its 
universal reconstruction capabilities for silver nanoclusters, where the 
network uncovers novel geometric structures that reproduce the experimental 
scattering data with very high precision.
\end{abstract}
\maketitle
	
\section{Introduction}
The imaging of systems of nanometer size is of great importance for many 
branches in biological, chemical and physical sciences. The laws of wave optics 
demand the usage of wavelengths in the x-ray regime. However, the large energy 
carried by each photon rapidly damages such delicate 
samples \cite{neutze2000}. The deterioration of the sample during the imaging 
process can be avoided if the sample image is generated on a much shorter 
timescale than that on which the destruction process, e.g. Coulomb explosion 
\cite{jurek2004}, occurs. This requirement is fulfilled by imaging using 
high-intensity ultra-short femtosecond pulses, as produced by free electron 
lasers \cite{Chapman2006,Gaffney2007}. Since the object's features and the 
wavelength are comparable, the resulting image is dominated by scattering 
features and, in order to reveal the underlying real-space image, 
further processing is necessary \cite{Chapman2006}. To date, 
improvements in object reconstruction allowed the investigation of ever smaller 
unsupported nanosystems such as viruses \cite{Seibert2011, Ekeberg_2015, 
Ayyer_2019}, helium droplets \cite{Gomez2014, Rupp_2017, Langbehn_2018}, 
rare-gas clusters \cite{Rupp2012}, or metallic nanoparticles \cite{Barke_2015}.

For very short wavelengths, i.e. hard x-rays, the scattering occurs 
predominantly at small angles. In this case, the scattering process can be 
understood in the Fraunhofer limit, and the scattering field is the 
two-dimensional Fourier transform of the projected electron density. 
A subsequent iterative phase retrieval then allows to reconstruct this 
two-dimensional density projection with high fidelity from a single scattering 
pattern \cite{marchesini2003, Seibert2011}. Further, individual 
scattering images of an ensemble of identical objects can be merged to obtain 
the three-dimensional object density \cite{Ekeberg_2015,Ayyer_2019,Ayyer_2021}. 
For non-reproducible targets, such tomographic techniques cannot be employed 
as only a single scattering image is available. In this situation, 
three-dimensional information can be extracted from wide-angle reflexes of the 
scattering pattern \cite{Raines2010}, which require longer wavelengths. Recent 
theoretical works indicate in principle the completeness of such three-dimensional 
information encoded in wide-angle scattering signals 
\cite{engel2020identity,engel2020modulus,engel2020reconstruction} 
for solid convex objects under narrow assumptions within the first Born 
approximation and infinitely many exact measurements. 
Yet, they pose a significantly more
complicated inversion problem compared to the small-angle reconstruction method
\cite{Raines2010,Barke_2015,Rupp_2017}. 
Further, in reality, the first Born approximation is usually insufficient, and 
processes such as absorption and re-scattering have to be included. Moreover, an experiment 
always has to manage with a finite number of measurement data.
The non-reproducibility of the particles further hinders the independent acquisition of additional
shape information using alternative experimental techniques. Hence, the reconstruction 
problem of single-shot wide-angle scattering is to find a particle reproducing the input
scattering pattern within the angular range of the detector, whilst obeying additional structural
constraints on the object, that may not necessarily result in a unique solution, given the 
experimental conditions \cite{Raines2010,wang2011non,wei2011fundamental}.
Thus far, these reconstructions mostly 
rely on iterative forward fitting methods that are based on simulations
of the scattering process of a suitably parametrized object model 
\cite{Barke_2015,Rupp_2017,Langbehn_2018}. While highly successful, the repeated 
scattering simulations are computationally expensive and are restricted to the 
assumed object model. 

Recent years have seen rapid development in image processing and reconstruction 
techniques based on deep learning methods 
\cite{hinton2006,lecun2015,goodfellow_DeepLearning}. These concepts have 
already found broad applications in statistical physics, particle and 
accelerator physics \cite{carleo2019review,baldi2014,kasieczka2017,kasieczka2019,meinert2020}, 
material sciences \cite{carleo2019review,laanait2019,laanait2019exascale,chen2021},
as well as for approximating solutions to differential equations 
\cite{raissi2019,raissi2020}. 
In diffractive imaging, deep learning techniques have been explored for 
the efficient reconstruction of both small-angle and wide-angle images. 
Phase retrieval and subsequent Fourier inversion with convolutional neural 
networks has been demonstrated for simulated small-angle scattering patterns 
\cite{Cherukara_2018}, and have been expanded to three dimensions for the
reconstruction of object densities from complete Fourier volumes 
\cite{chan2020}. On the experimental side, the pre-selection of automatically 
recorded scattering patterns into various categories has been implemented as 
a classification task \cite{Langbehn_2018}, and generative learning helped 
to reveal common features in patterns connected to object classes and 
imaging artifacts \cite{Zimmermann2019}. 
Recently, shape and orientation of icosahedral silver nanoclusters were 
reconstructed from experimental wide-angle scattering patterns using a neural 
network trained solely on simulated training data \cite{stielow2020}.
This was achieved by utilizing a convolutional neural network that, combined 
with data augmentation techniques, is capable of processing experimental images 
that suffer from a variety of physically relevant artifacts and defects.

In this article, we present a neural network approach for reconstructing shape 
and orientation of arbitrary nanoclusters from single-shot wide-angle 
scattering images that does not depend on the parametrization of the object 
model. Instead, we use a voxel model of the object density similar to that used 
in small-angle scattering \cite{chan2020}. For that, an encoder-decoder 
architecture is employed that realizes the transition from the two-dimensional 
image to the three-dimensional object space. The interpolation beyond the 
underlying training data set is improved by implementing physics-informed 
learning, in which the theoretical scattering model itself is included in the 
loss function. 

The article is organized as follows. In Sec.~\ref{sec:dataset}, 
we briefly review the scattering simulation method that is based on the 
multi-slice Fourier transform (MSFT) algorithm, and we introduce the 
construction of the basis set and its augmentations. The design of the neural 
network including the physics-informed training scheme is presented in 
Sec.~\ref{sec:network}. Its capabilities and limits are discussed in 
Sec.~\ref{sec:analysis}, followed by the evaluation of experimental data in 
Sec.~\ref{sec:experiment} and some concluding remarks in Sec.~\ref{sec:summary}.

\section{Modelling and simulating scattering of silver nanoclusters}
\label{sec:dataset}

Scattering experiments with light in the x-ray regime are known to reveal 
structure information such as geometric shapes, spatial orientation and size of 
nanoparticles, in some cases also their internal structure 
\cite{Ekeberg_2015,Sander2018}. 
Here, we focus on the reconstruction of silver nanoparticles that had been 
illuminated with soft x-rays from a free electron laser with wavelength 
$\lambda=13.5$nm. At this wavelength, scattering off these clusters with sizes 
between $50...400$nm can then be regarded as in the wide-angle limit. The 
nanoparticles are produced by a magnetron sputtering source in a cluster beam 
machine. The generated stream of nanoclusters shows a wide 
range of shapes and sizes, meaning that the particle shapes occur to a certain 
extent randomly. Moreover, each individual experiment is non-reproducible as 
the Coulomb explosion prevents multiple illumination. It is also known that the 
particles emerging from the source have not yet relaxed to an equilibrium 
state at the time of illumination, hence geometric structures such as 
icosahedra have been found \cite{Barke_2015,stielow2020} that are not expected 
to be stable for large particle sizes.

Due to the lack of a direct inversion algorithm for the reconstruction of 
geometric information from a single-shot wide-angle scattering image, 
comparative methods such as forward fitting have been employed 
\cite{Barke_2015,Sander2015,Langbehn_2018}. The theoretical scattering patterns 
are generated using a multi-slice Fourier transform (MSFT) algorithm that takes 
absorption into account but neglects multiple scattering events as well as 
momentum transfer to the nanoparticle. Because of the short absorption length 
of $12.5$nm in silver, this algorithm gives very accurate results. Most 
importantly, it can be represented as a linear tensor operation which makes it 
suitable for efficient parallel computation.

For an efficient implementation of a reconstruction algorithm, a suitable 
parametrization of the object is needed. Typically, this means a restriction of 
the class of object shapes to a finite set of highly symmetric base solids
with relatively few degrees of freedom. For nanoparticles out of equilibrium, 
however, transient shapes need not necessarily be highly symmetric. This in 
turn implies a trade-off between reconstruction accuracy and numerical 
efficiency. Already in the case of only few parameters, neural networks 
outperform conventional forward fitting based on Monte Carlo simplex methods 
\cite{stielow2020}, which is expected to become even more prominent with 
increasing number of degrees of freedom. The limiting case is to 
represent the object on a discrete three-dimensional grid; such representations 
are commonly used for the reconstruction of real-space objects from a series of 
images using deep neural networks \cite{choy2016}. In the realm of scattering 
physics, this representation has been employed for the reconstruction of a 
reproducible nanoparticle from a three-dimensional scattering pattern that has 
been compiled from a series of small-angle scattering images \cite{chan2020}.
We show here that the discretized three-dimensional object can be reconstructed 
from a single wide-angle scattering pattern using deep neural networks.

\subsection{Object classes for training the neural network}

\begin{figure}[tb]
\includegraphics[width=\columnwidth]{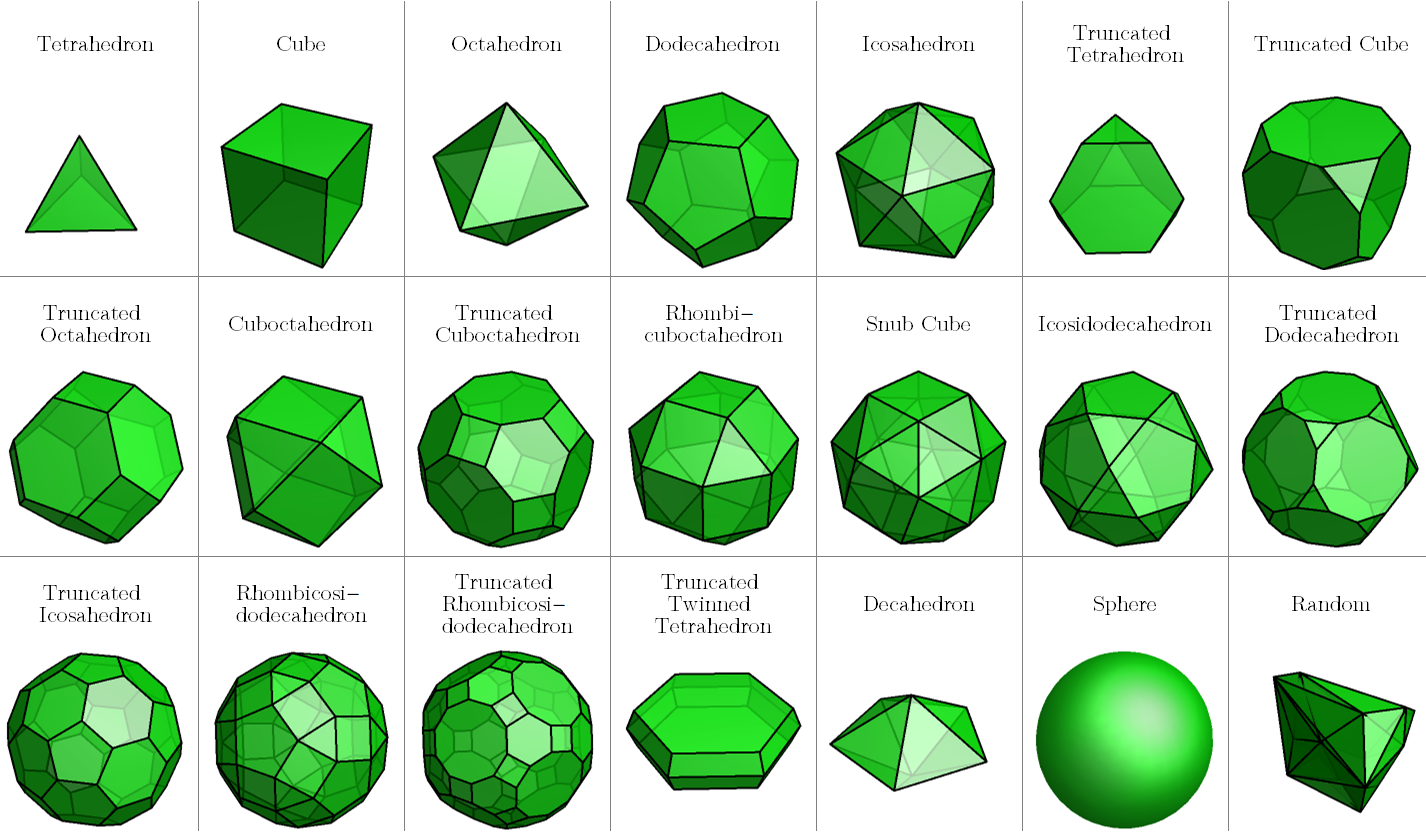}
\caption{The basis set of 21 shapes contains all Platonic and Archimedean 
solids (except for the snub dodecahedron) and, additionally, the decahedron, 
the truncated twinned tetrahedron, spheres and polyhedra with fully randomized 
vertices, defined by enclosing 50 random space points.}\label{fig:baseSolids}
\end{figure}

The training of a neural network requires a suitably chosen set of training 
data. 
Due to the large number of atoms in a nanocluster (typically on the order of $10^9$), 
the silver nanoparticles can be assumed to be macroscopic dielectric bodies, that are well described 
by a binary permittivity function 
$\epsilon_r(\mathbf r) = \epsilon_\text{silver}$ for $\mathbf r \in V_\text{object}$, 
and $\epsilon_r = 1$ otherwise \cite{Barke_2015}. Additionally, we demand all objects to be convex,
as this a necessary condition for the existence of a unique solution in the ideal scenario of
informationally complete measurements \cite{engel2020modulus}.
In order to account for a large variety of (convex) object shapes that 
still contain some symmetry, we choose a basis set that contains all Platonic 
solids, all Archimedean solids (except the snub dodecahedron), the decahedron 
and truncated twinned tetrahedron, as well as spheres and convex polyhedra with 
fully random vertices. This set is depicted in Fig.~\ref{fig:baseSolids}. 
Further, these base solids have been stretched and squashed along one of their 
symmetry axes, and have been randomly scaled and rotated for maximum 
flexibility. Despite the still finite number of objects, it is expected that a 
large enough portion of object space is covered, and that the neural network 
is capable of interpolating efficiently between them. Note, however, that some 
of the included objects (such as the tetrahedron) are highly unlikely to ever 
occur in an experiment but are included nonetheless.

\subsection{Scattering simulation}
The training data are obtained numerically by employing the MSFT scattering 
framework. All objects have been rasterized on a three-dimensional grid of 
$192 \times 192 \times 192$ points and are stored as flattened png images. For 
each object, the corresponding scattering intensity pattern is calculated using
the MSFT algorithm. The lateral dimensions of the object are padded to $512 
\times 512$ pixels upon simulation, and the resulting real transfer momentum 
space covers $128 \times 128$ pixels. As the transverse intensity decreases 
exponentially away from the image center, the intensity values are scaled 
logarithmically in order to preserve important scattering features at large 
transfer momenta. In addition, in order to simulate detector dark noise, a 
random constant offset is being applied before scaling. Each image is then 
normalized and stored as a png image. As the object rasterization as well as 
the MSFT scattering calculations require considerable computation 
times, a data set of 140\,000 objects has been pre-generated and stored.

\onecolumngrid

\clearpage
\twocolumngrid

\subsection{Simulating experimental artifacts by image augmentation}

The theoretical training data contains the maximal amount of information 
regarding the light scattering off a nanoparticle allowed by scattering and 
detection physics. However, in experimental situations, technical limitations 
often obscure some of the information necessary to, e.g. identify the shape of 
a particle. For example, all images contain a central hole that protects the 
detector from the central unscattered beam. This is such a prominent artifact
that a neural network is very likely to regard this as the most important 
feature, whereas the information about the shape of the particle resides in the 
outer fringes of the scattering pattern. Therefore, such defects have to be 
included in the training of the network from the outset. 

In Ref.~\cite{stielow2020} it was demonstrated that data augmentation 
techniques can be used to simulate these measurement artifacts and to train a 
neural network that is robust against such effects. We extend this augmentation 
approach by introducing additional filters and on-the-fly augmentation. Rather 
than pre-generating a set of augmented images, here we apply random 
augmentations at each training step. Hence, every time the network is presented 
with the same data point, a random augmentation filter is being selected, which 
helps to prevent overfitting. 

\begin{figure}[htb]
\includegraphics[width=\columnwidth]{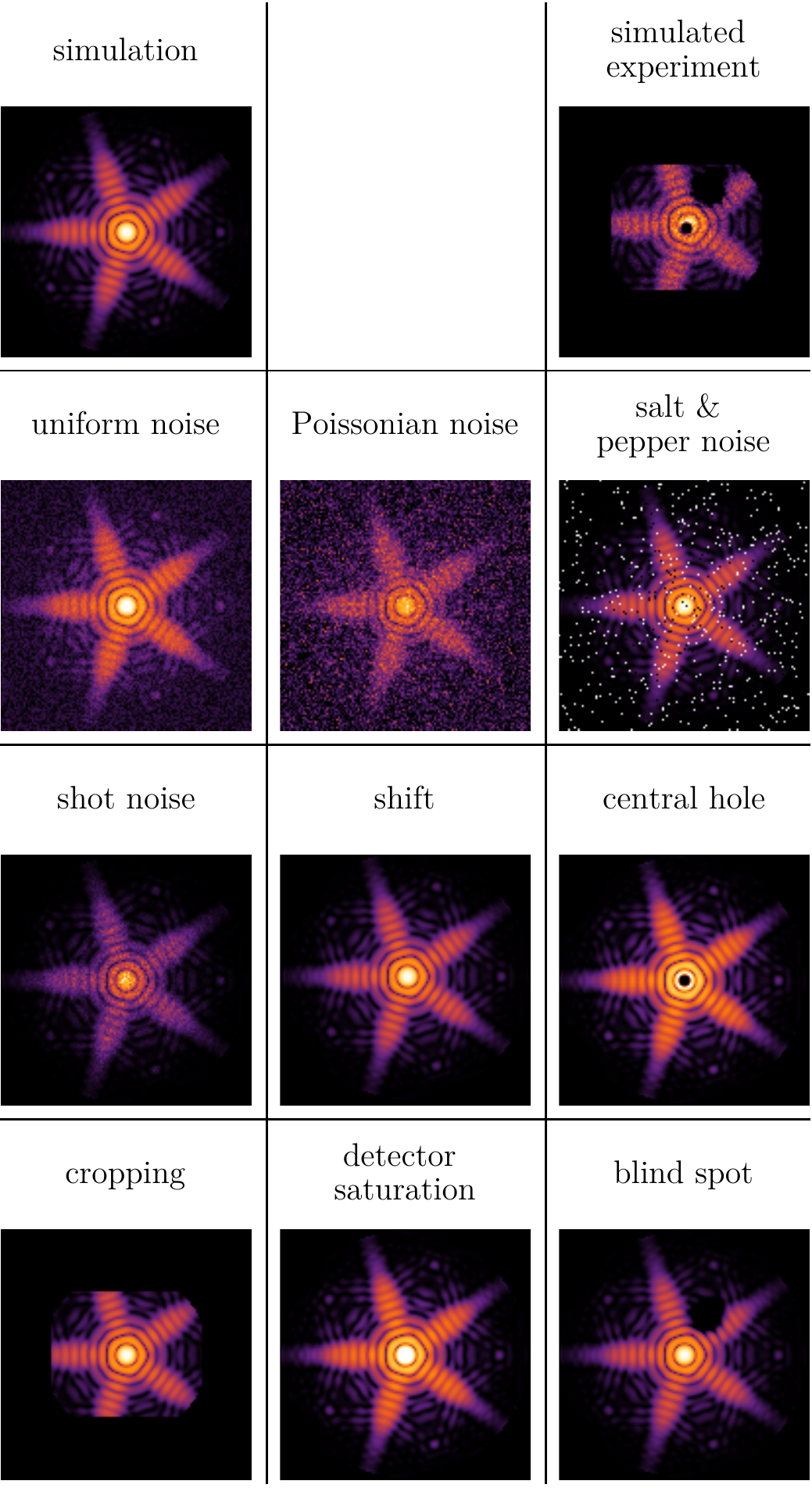}
\caption{Image augmentation is used to feed the neural network scattering patterns
with various defects to increase its prediction robustness. Each simulated 
scattering pattern (top left) is modified with one of the nine fundamental 
filters (bottom $3 \times 3$ square) or a combination of them (top right) to 
mimic experimentally obtained scattering patterns.}\label{fig:augmentations}
\end{figure}

Examples of all used augmentation filters are shown in 
Fig.~\ref{fig:augmentations}. The augmentation functions \textit{uniform 
noise}, \textit{salt \& pepper noise}, \textit{shift}, \textit{central hole} 
and \textit{blind spot} have been implemented as described in 
Ref.~\cite{stielow2020}. The \textit{cropping} filter has been modified to 
simultaneously apply rectangular and circular cropping masks with random 
sizes. The \textit{Poissonian noise} filter has been implemented by adding a 
random matrix sampled from a Poissonian distribution with variance $\lambda = 
1.0$ to the normalized scattering pattern, while the \textit{shot noise} filter 
multiplies the scattering pattern with a random Poissonian matrix with variance 
$\lambda = 10^{r + 1}$ where $r$ is an uniform random number from the interval 
$[0,1]$. These filters account for the Poissonian background counts as well as 
the discrete nature of photons in the low-intensity limit. The \textit{simulated 
experiment} filter is implemented by a consecutive application of the
\textit{shot noise}, \textit{shift}, \textit{blind spot}, \textit{detector saturation}, 
\textit{central hole}, \textit{cropping}, and \textit{shift} filters.

\section{Design and training of the scattering reconstruction network}
\label{sec:network}

In classical image processing, the task of creating a three-dimensional model 
from one or more two-dimensional images is a well-known problem that can be 
efficiently tackled using neural networks \cite{niu2018,choy2016}. The 
reconstruction of a discretized three-dimensional object from a two-dimensional 
single-channel image requires a dimension conversion, which is commonly solved 
with encoder-decoder architectures. 
In this case, the input image is projected into a latent space from which the 
conversion into the output space is performed. When implementing multi-view 
reconstructions of macroscopic objects from photographic images, additional 
recurrent elements within the latent space are required \cite{choy2016}.

\begin{figure}[htb]
\includegraphics[width=\columnwidth]{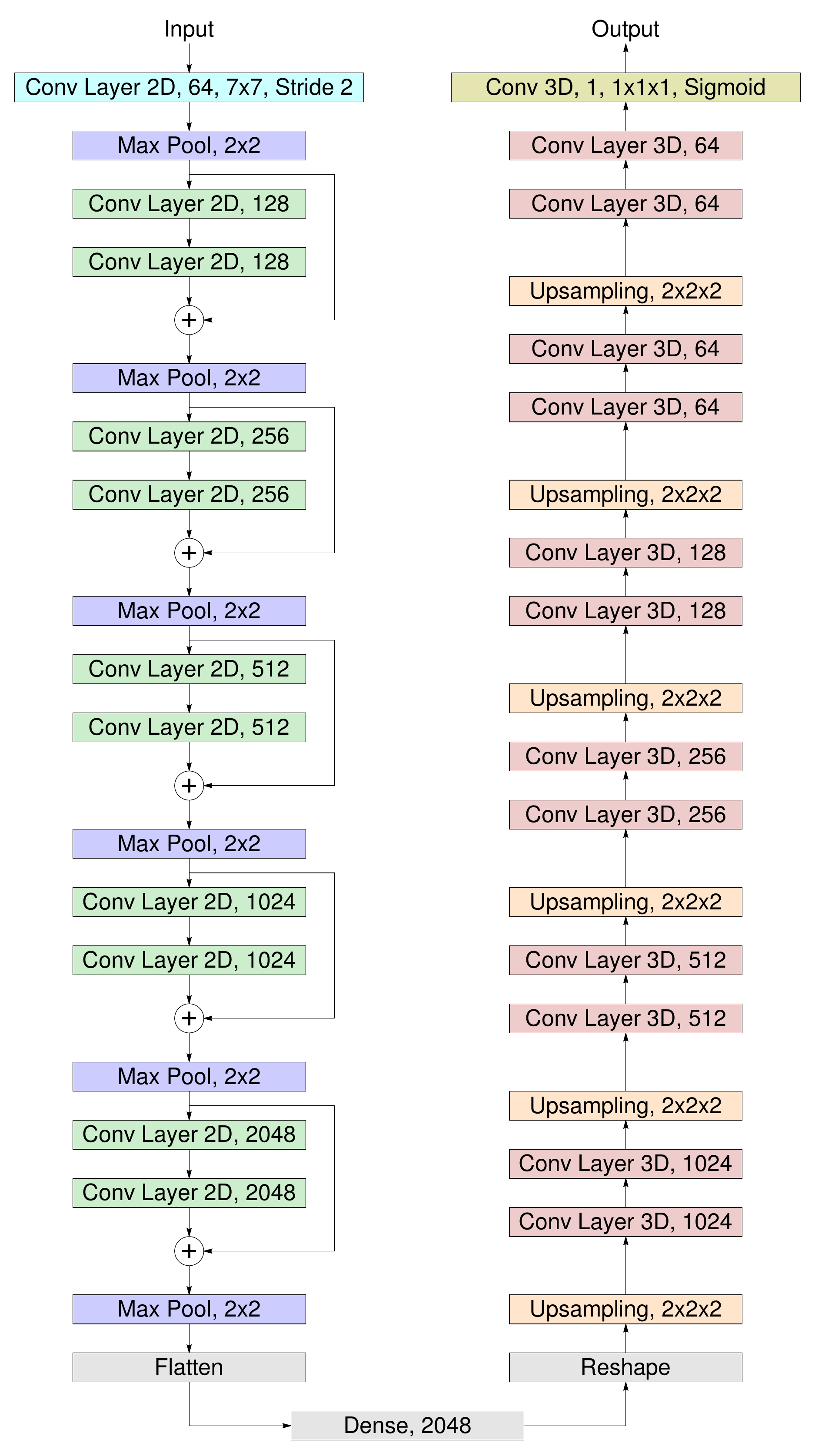}
\caption{Neural network with encoder-decoder structure. The encoder (left 
column) consists of five residual blocks each containing two consecutive 2D 
convolution layers with $3 \times 3$ kernels. The filter size is doubled
with each residual block, while the lateral dimensions are reduced by pooling layers. 
The latent space (bottom) is one-dimensional and is further connected by a dense layer. 
After reshaping, the decoder (right column) applies $2 \times 2 \times 2$ 
upsampling operations followed by two 3D convolution layers each.
All convolution layers are regularized with a dropout ratio of 0.2 and batch normalization 
is applied before the leaky ReLU activation.}\label{fig:network}
\end{figure}	

The architecture we developed for single-shot scattering reconstructions is 
depicted in Fig.~\ref{fig:network}. The encoder section of the network in the 
left column is constructed as a residual convolutional lateral compressor. 
An initial pickup layer with 7 $\times$ 7 convolution kernels and stride 
2, followed by Max pooling operations, is used to rapidly convert the input 
tensor size from $128 \times 128 \times 1$  to $32 \times 32 \times 64$
elements. Following that is a sequence of five residual blocks, each halving 
the lateral size further while doubling the number of filters. Every residual 
block consists of two consecutive convolution layers as well as an identity 
shortcut which are combined by a summation layer \cite{He_2016resNet}. Each 
convolution layer has a kernel size of $3\times 3$ and is activated by the leaky 
ReLU function 
\begin{equation}
\mathrm{lReLU}(x) =  \left\{ \begin{array}{ll} x& \text{if } x > 0 \, 
\text{,}\\ 0.01x & \text{otherwise.} \end{array}\right.
\end{equation}
after regularization by batch normalization and dropout. Within the latent 
space, an additional fully connected layer with 2048 neurons is employed. The 
decoder (right column of Fig.~\ref{fig:network}) is designed in reverse with 
upsampling layers instead of pooling and three-dimensional convolution layers. 
Unlike the encoder, the decoder does not employ residual operations and is 
instead of linear structure, as residual connections were found to offer no 
improvement in the prediction quality while increasing the training time 
significantly. The final compression of the filter dimension into the output 
tensor of size $64 \times 64 \times 64 \times 1$ is performed by a 
three-dimensional convolution operation with a $1 \times 1 \times 1$ kernel 
and sigmoid activation, as the output tensor is of binary character. The full 
network has now approximately 200 million free parameters.

\subsection{Physics-Informed Learning}\label{sec:physInf}

\begin{figure}[htb]
\includegraphics[width=\columnwidth]{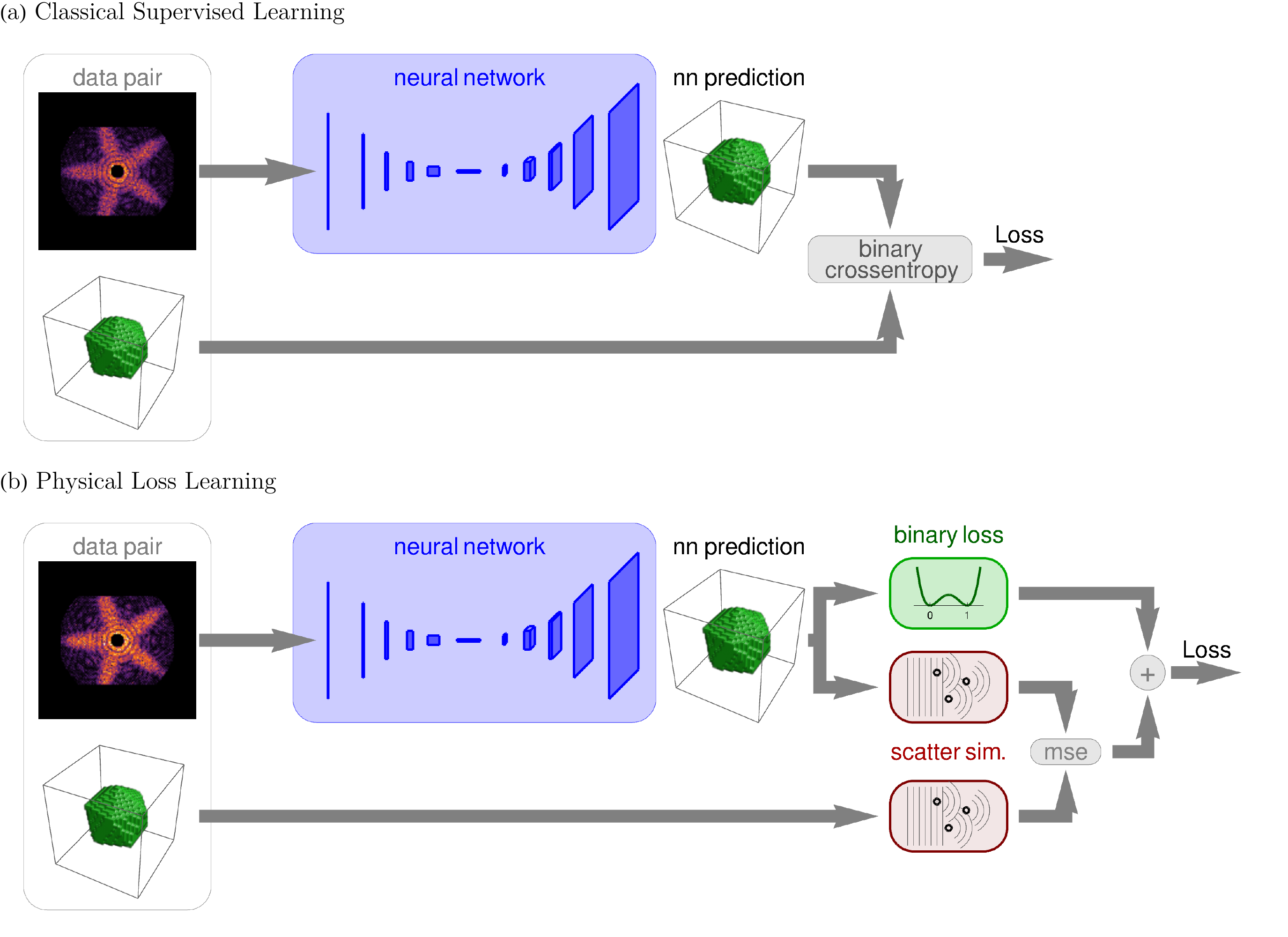}
\caption{In classical supervised learning (a), the loss score is determined by 
the binary crossentropy between the network prediction and the target entry of 
each data pair. In the physical learning scheme (b), the loss score is 
calculated within the scatter space rather than the object space. This is 
done by simulating the scattering pattern of both the network prediction 
as well as the target object, and calculating their mean squared difference 
(scatter loss). To enforce the binary nature of the object model, an 
additional regularization function (binary loss) is applied to the 
prediction.}\label{fig:physLossScheme}
\end{figure}

Classical supervised learning consists of comparing the predictions $\mathbf{p}$ 
made by the neural network on the training inputs $\mathbf{x}$ to the corresponding 
ground truth targets $\mathbf{y}$, and calculating a loss score as illustrated 
in Fig.~\ref{fig:physLossScheme}(a). However, a straightforward implementation 
of this idea is unfeasible in our situation. Silver has a rather short 
absorption length of $12.5 \,\text{nm}$ at the relevant photon energies, which 
is much shorter than the cluster diameters that range from 63 to 320~nm. As a 
result, the incoming radiation does not penetrate the entire nanoparticle and, 
in particular, has no access to those parts of the scattering object that are 
furthest away from the radiation source. This is turn means that a significant 
part of the object does not contribute to the scattering image. However, the 
penalizing loss function forces the neural network to attempt to reconstruct 
those regions for which very little information is contained in the input image.
Hence, the neural network is either forced to complete the object from 
symmetric projections (which is indeed observed to some degree), or is driven 
into significant overfitting. This in return leads to poor generalization
capabilities on shapes outside the training data set, an example of which is provided
in the Appendix.

In order to ensure that the neural network learns only from physically relevant 
information, we propose the calculation of a loss score in scattered space, 
which is shown in Fig.~\ref{fig:physLossScheme}(b). Instead of comparing the 
prediction $\mathbf{p}$ with the target $\mathbf{y}$ directly by the mean 
binary crossentropy
\begin{align}
H(\mathbf{y},\mathbf{p}) = &\frac{1}{N^3}\sum_{i,j,k=1}^N  \Big[ y_{i,j,k} \, 
\log(p_{i,j,k})
\nonumber\\
& + (1-y_{i,j,k})\,\log(1-p_{i,j,k})\Big] \, ,
\end{align}
both $\mathbf{p}$ and $\mathbf{y}$ are used as inputs for the MSFT algorithm,
and the loss is calculated by the mean squared distance of the resulting 
scattering patterns, scaled logarithmically. This so called scatter loss can be 
expressed as
\begin{align}
L_\mathrm{s}(\mathbf{y},\mathbf{p})= &\frac{1}{M^2}\sum_{i,j=1}^M 
\left[\log\left(\left|{\mathbf{E}_\text{MSFT}(\mathbf{y})}_{i,j}\right|^2 + 
\epsilon\right) \right. \nonumber \\
&-\left.\log\left(\left|{\mathbf{E}_\text{MSFT}(\mathbf{p})}_{i,j}
\right|^2 + \epsilon\right) \right]^2 \,,
\label{eq:scatterloss}
\end{align}    
with some chosen noise level $\epsilon$, and where $\mathbf{E}_\text{MSFT}$ is 
the normalized electric-field distribution obtained by the MSFT algorithm.
In this way, the training goal of the neural network is moved from predicting 
the real-space shape of an object to generating an object volume that 
reproduces the input scattering pattern.

Although the terminal layer of the neural network is sigmoid activated, this 
activation does not enforce the binary nature of our particle model. Therefore, 
we introduce an additional regularization term to the loss function 
\eqref{eq:scatterloss} by penalizing non-binary object voxels with the binary 
loss function 
\begin{equation}
L_\mathrm{b}(\mathbf{y},\mathbf{p})= \frac{1}{N^3}\sum_{i,j,k=1}^N (p_{i,j,k})^2 \, (1 - 
p_{i,j,k})^2 \,.
\label{eq:binaryloss}
\end{equation}
The binary loss function \eqref{eq:binaryloss} is weighted by a factor $0.1$ 
compared to the scatter loss \eqref{eq:scatterloss} to ensure optimal 
convergence. This is an instance of physics-informed learning 
\cite{raissi2019,raissi2020} where physical laws are incorporated in the 
training function.

\onecolumngrid

\clearpage
\twocolumngrid

\subsection{Network Training}

\begin{figure}[htb]
\includegraphics[width=\columnwidth]{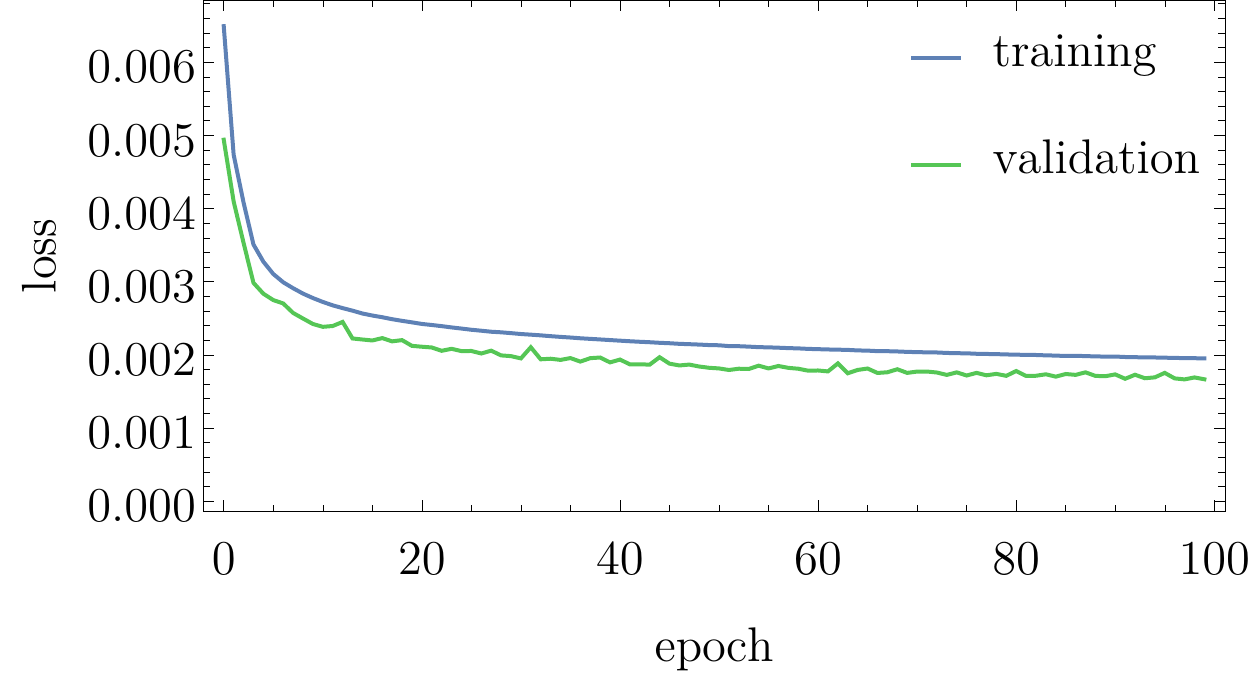}
\caption{The training loss of the neural network converges within 50 full 
cycles of the training set to a near halt. The loss on the validation set 
follows a similar trajectory, but is consistently smaller than the training 
loss, due to the absence of augmentations and 
regularization.}\label{fig:learningCurve}
\end{figure}

The neural network was implemented and trained within the TensorFlow 2.3.1 
Keras framework and Python 3.6.6. The binary loss regularization and scatter 
loss were both implemented as TensorFlow functions, thereby enabling
backpropagation on GPU devices during training. We have chosen the adaptive 
moments (ADAM) gradient descent optimizer for optimal convergence. The training 
dataset was pre-generated, and scattering patterns were stored as png images,
while object densities were rescaled and saved as $64 \times 64 \times 64$
numpy arrays to minimize hardware access and processing times. 
The data set contains 140\,000 samples in total and has been split into a training 
and a validation set with a ratio $5:1$. The 
training set was re-shuffled before each epoch, and data was read from the hard 
drive and randomly augmented on-the-fly. The validation data was not augmented 
in order to monitor the peak reconstruction capability. Training was performed 
on a dedicated GPU server with two Intel Xeon Silver 4126 CPUs and four Nvidia 
RTX2080ti GPUs. Distribution of each training batch over 
all four GPUs allowed a maximum batch size of 32. We found the optimal training 
duration to be 50 epochs for sufficient convergence. The corresponding learning 
curve of the network used throughout this manuscript is shown in 
Fig.~\ref{fig:learningCurve}. The total training time accumulated to 63h. 
   
A consistent result over different training runs from independent random 
initializations could only be achieved by applying regularization in 
every layer. Batch normalization counteracts the tendency to no-object
predictions. Simultaneously, dropout regularization prevents the neural network 
from converging to non-physical predictions, which may produce similar 
scattering patterns but are non-binary point clouds in object space that do 
not correspond to solid convex (or at least star-shaped) bodies. The combined 
effect of these regularization is that the training loss in 
Fig.~\ref{fig:learningCurve} shows no overfitting compared to the validation 
loss. However, this cannot rule out the possibility of overfitting to either 
the underlying set of solids or the augmentations used.

\section{Prediction capability of the neural network}
\label{sec:analysis}

\begin{figure}[htb]
\includegraphics[width=\columnwidth]{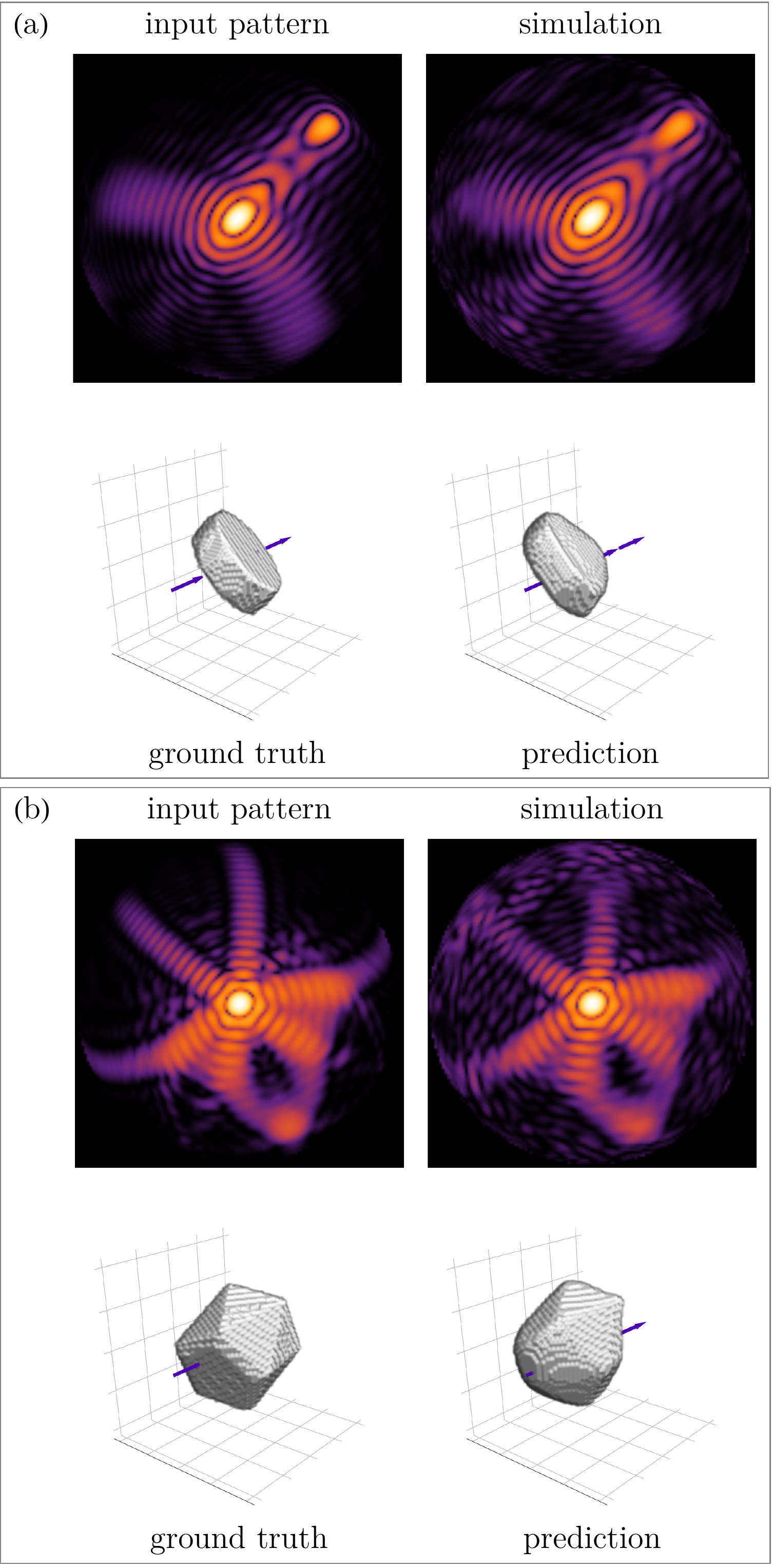}
\caption{Scattering patterns and real-space object shapes are reproduced by the 
neural network for most objects of the test set, such as the 
rhombicosidodecahedron (a). For some examples, the predicted object is 
reconstructed without the far side or sports a shallow dome in the beam 
direction (b), both of which have no significant impact on the
scattering pattern.}\label{fig:choppedSet}
\end{figure}

During training of the neural network, we benchmarked its prediction 
capabilities on the validation set which was generated from the same object 
space as the training set. In order to test its interpolating power, we created 
an additional test set of object data unknown to the network. These bodies
were created by truncating the previously scaled and stretched object classes 
along random symmetry axes, thus breaking some of the symmetries and creating 
new shapes. In this way, a total of 1000 new objects were created. 

In the majority of cases, the neural network is capable of detecting the new 
deformations. An example is shown in Fig.~\ref{fig:choppedSet}(a), 
corresponding to a heavily truncated rhombicosidodecahedron. The object 
prediction of the neural network (bottom right) closely resembles the ground 
truth of the object (bottom left), while their scattering patterns are nearly 
indistinguishable (top row in Fig.~\ref{fig:choppedSet}(a)).
This implies that, due to its physics-informed training, the neural network 
does not merely interpolate between known shapes, but rather composes an 
hitherto unknown object from facets associated with distinct reflexes in the 
scattering pattern. 

\begin{figure}[htb]
\includegraphics[width=\columnwidth]{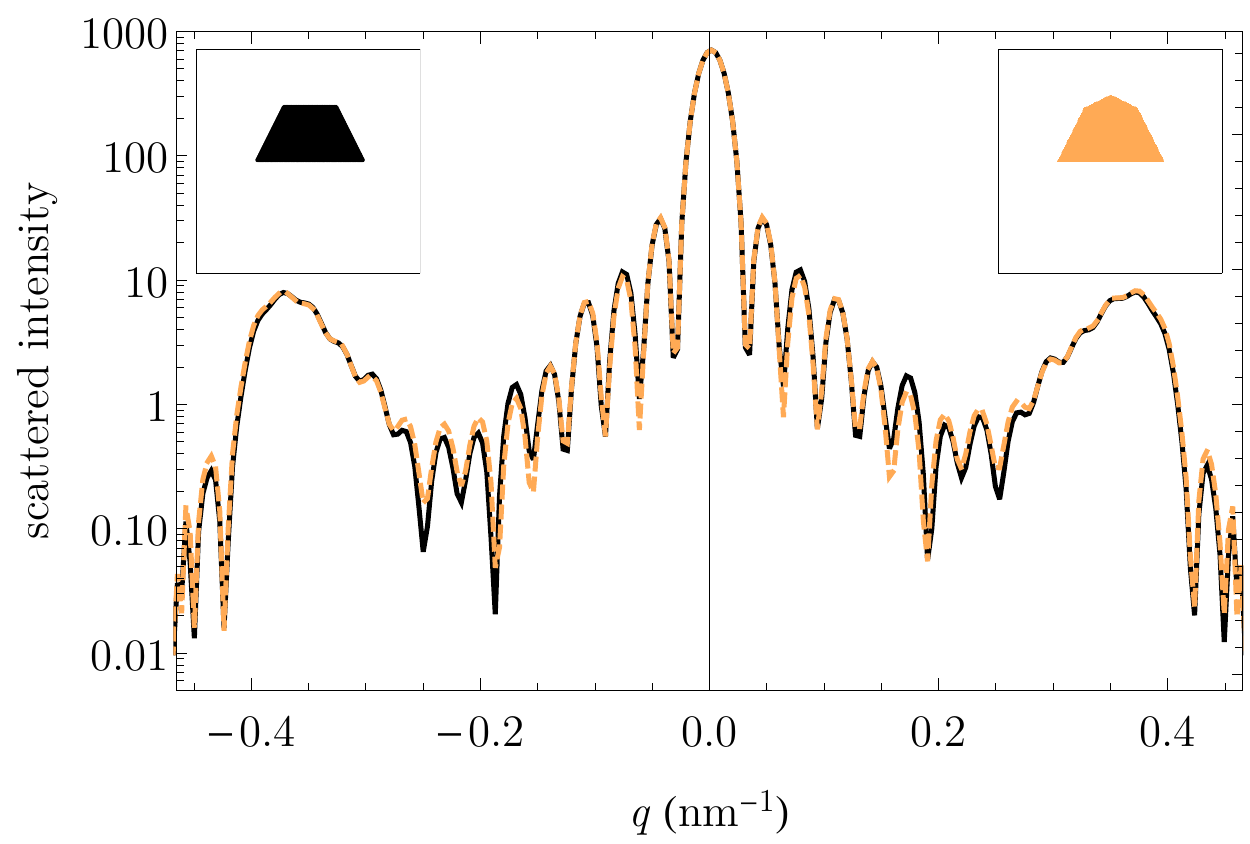}
\caption{The scattered intensity signals of a truncated triangle with a 
footprint of $212.5 \, \text{nm}$ and of the same object equipped with a 
shallow tip of $25\%$ of its height are almost identical.}\label{fig:domeing}
\end{figure}

Conversely, this also implies that objects are only constructed from real-space 
features that impact the scattering pattern. An example is shown in 
Fig.~\ref{fig:choppedSet}(b), where two significant effects can be observed. 
First, the far side of the predicted object (bottom right) is featureless. This 
was expected because of the strong absorption of the incoming radiation which 
prevents a significant contribution from the scattering off these regions. The 
same effect was also observed on the validation set and even the training set.
The neural network then either cuts off the far side completely, or replaces it 
with a smooth droplet shape.
Second, the flat front facet of the input object (bottom left) is being 
converted into a shallow dome. Surfaces oriented close to perpendicular with 
respect to the incoming beam are particularly difficult to reconstruct, as the 
strongest associated reflexes appear in the backscattering direction. These 
reflexes would only be observable in a $4\pi$ detector configuration, for which 
the MSFT algorithm does not give reliable results. A simplified two-dimensional 
model of this effect is shown in Fig.~\ref{fig:domeing}, where a triangular 
shaped dome (orange object) is being added to a flat facet of a trapezoidal 
base (black object). The corresponding one-dimensional scattering intensity 
profiles are almost indistinguishable, in particular given a finite detector 
resolution.

\begin{figure}[htb]
\includegraphics[width=\columnwidth]{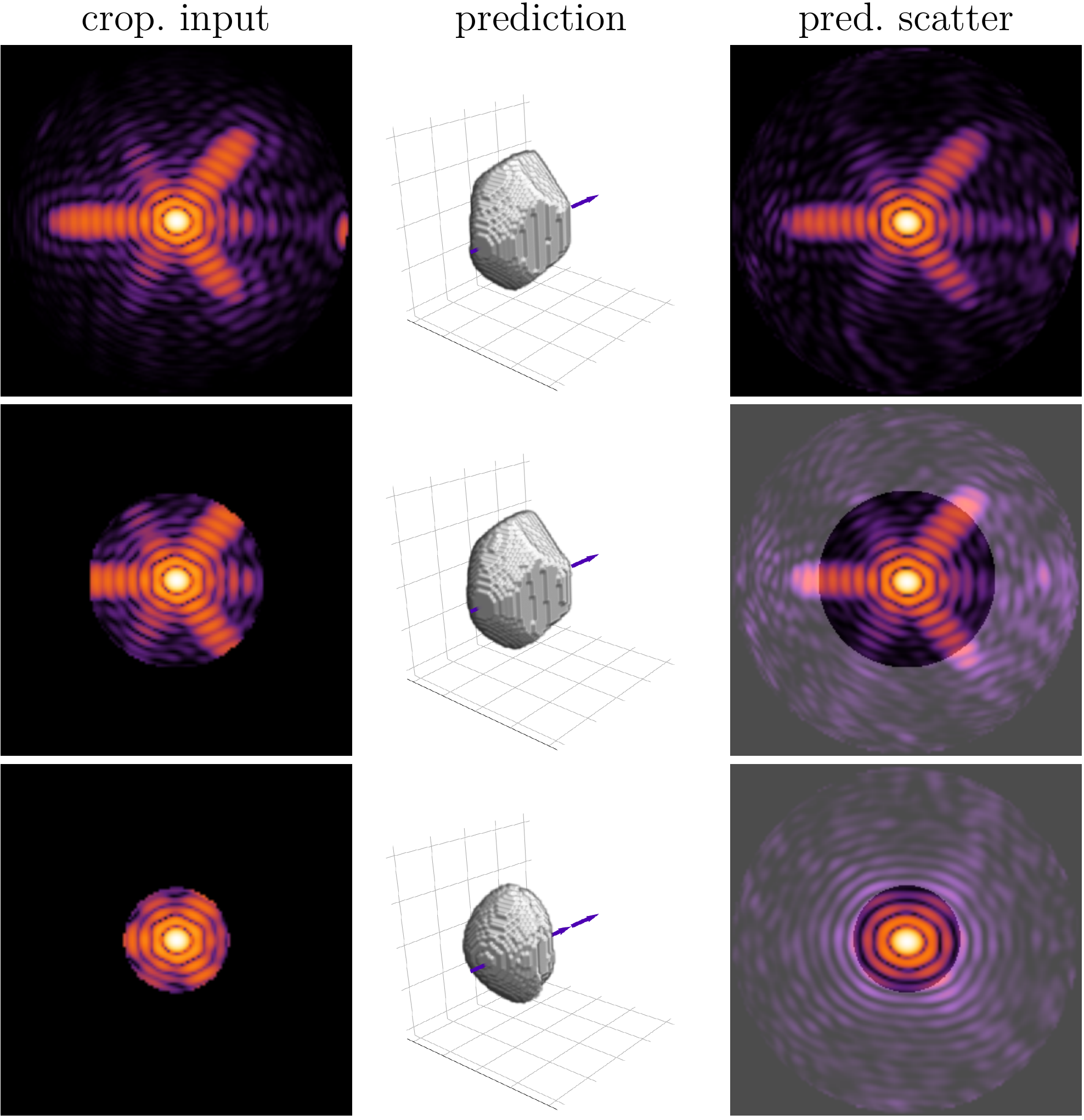}
\caption{Shrinking the angular span of the detection range (left column) 
leads to the loss of high-frequency information in the scattering pattern. 
Thus, the neural network predictions (central column) appear less crisp,
and corners and edges are rounded, while the corresponding scatter simulation 
(right column) still matches the input pattern within the input 
region (framed by gray mask).}\label{fig:croppingEffect}
\end{figure}

Delicate features of the real-space object appear at large transverse transfer 
momentum, that is, at large detection angles. During augmentation, this region 
is quite often cropped, giving the neural network the incentive to gather its 
information from the inner regions of small transfer momentum. This restriction 
is motivated by the limited detection angle of typical experiments. In order to 
understand the effect of cropping, we show in Fig.~\ref{fig:croppingEffect} the 
reconstructed images from the same input data pair for a series of ever smaller 
detection angles. As expected, with smaller available transfer momenta, the 
reconstruction quality decreases because information on sharp features is lost. 
As a consequence, edges and corners appear smoothed, while the facets are still 
recognizable.

\section{Neural network reconstruction of experimental data}
\label{sec:experiment}

So far, the neural network has been tested on synthetic data that capture the 
relevant scattering physics, and that have been augmented in order to mimic 
expected experimental artifacts. The trained network is now being used to 
reconstruct experimental single-shot wide-angle scattering data of silver 
nanoclusters \cite{Barke_2015}. Our choice has been informed by the existence 
of classical reconstructions using forward fitting methods with parametrized 
polyhedra, which provides the opportunity for direct comparison between the 
methods.

\begin{figure}[htb]       
\includegraphics[width=\columnwidth]{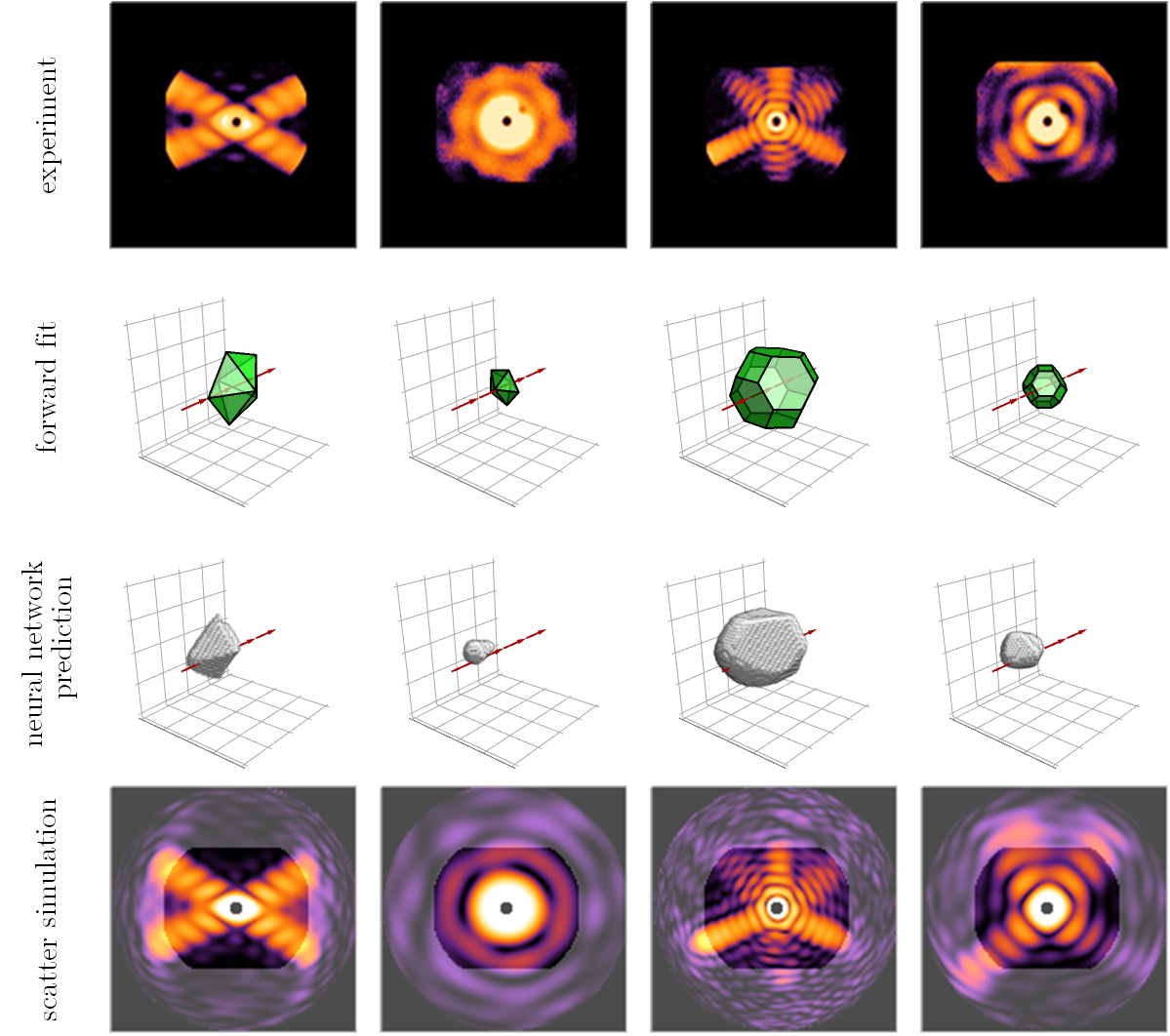}
\caption{The neural network is tested with the first half of the experimental scattering patterns 
from Ref.~\cite{Barke_2015} (left top row, permitted by Creative Commons CC-BY 
4.0 license (\url{http://creativecommons.org/licenses/by/4.0/)}) and the 
corresponding shape candidates obtained by forward fitting (second row, green solids). The 
neural network predictions are shown in gray in the third row. The simulated scattering patterns 
(bottom row) show excellent agreement with the input pattern inside 
the available region (confined by the gray masks).}\label{fig:expPred1}
\end{figure}

\begin{figure}[htb]       
\includegraphics[width=\columnwidth]{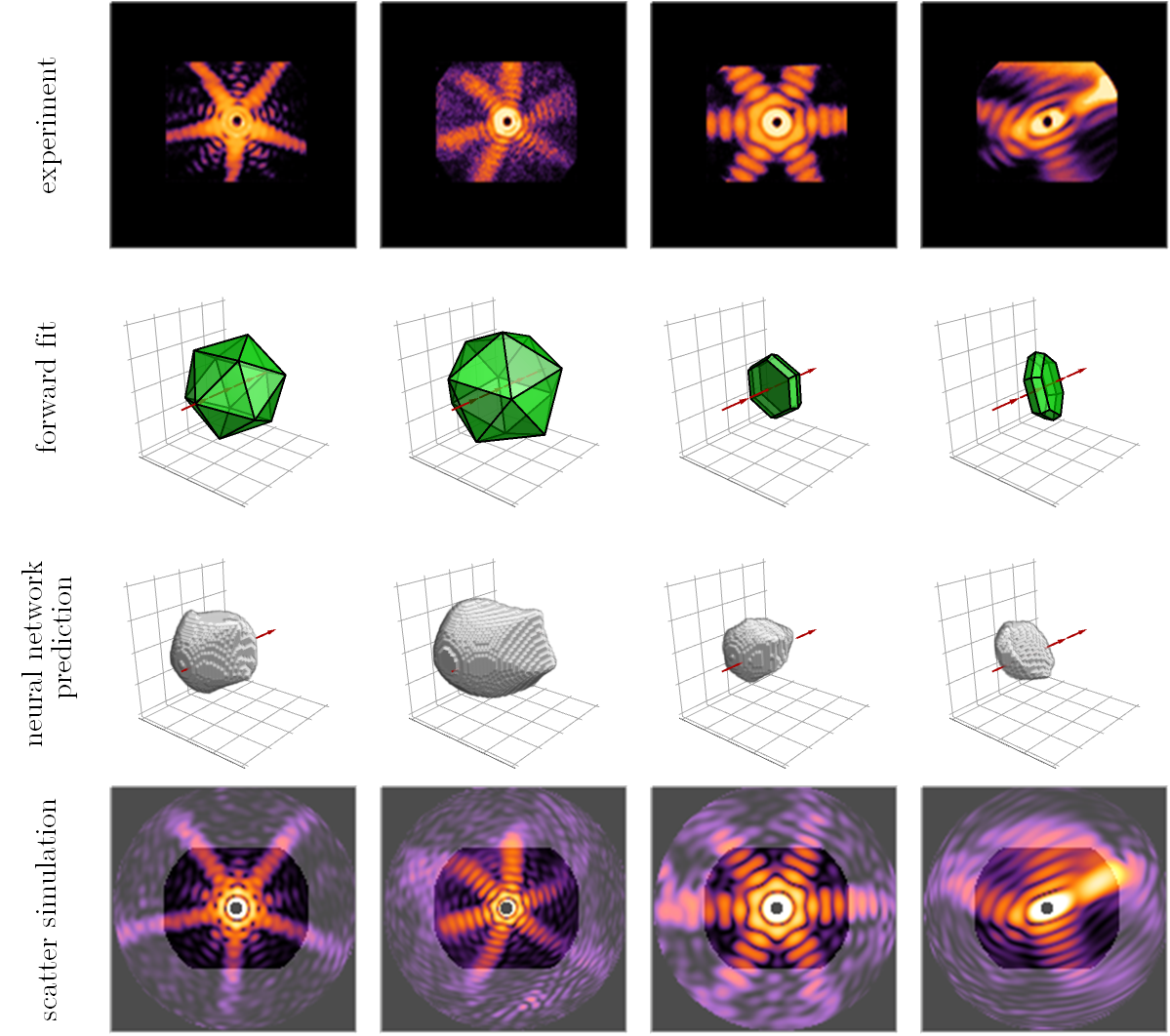}
\caption{The neural network is tested with the second half of the experimental scattering patterns 
from Ref.~\cite{Barke_2015} (left top row, permitted by Creative Commons CC-BY 
4.0 license (\url{http://creativecommons.org/licenses/by/4.0/)}) and the 
corresponding shape candidates obtained by forward fitting (second row, green solids). The 
neural network predictions are shown in gray in the third row. The simulated scattering patterns 
(bottom row) show excellent agreement with the input pattern inside 
the available region (confined by the gray masks).}\label{fig:expPred2}
\end{figure}

In Figs.~\ref{fig:expPred1} and \ref{fig:expPred2}, 
we compare the reconstructed nanoclusters from both 
the forward fitting (green objects in central column) and the neural network 
(grey objects in central column). The left column contains the experimental 
data from Ref.~\cite{Barke_2015}, whereas the right column depicts the 
simulated scattering profiles of the neural network predictions. We have 
explicitly shown the detection area to indicate the region which the neural 
network aims to reproduce. As discussed above, due to the lack of available 
large transfer momenta, the reconstructed objects by the neural network have 
smoother edges and corners. In comparison, the forward fit assumes the 
existence of sharp features which is unsupported given only the available 
information.
Also, as expected from the above discussion, the far sides of the reconstructed 
objects are either missing or being replaced by a smooth droplet, and shallow 
domes appear on their fronts.

\onecolumngrid

\twocolumngrid

Notwithstanding, the main facets are being reconstructed reliably, resulting in 
structures with globally similar features. However, the neural network predicts 
more elongated bodies which reproduce the softer interference patterns in the 
scattering reflexes. Moreover, the reconstructed bodies are no longer perfectly 
symmetric as assumed in the parametrized model, but show local defects that 
break certain symmetries. Note that the experimental scattering patterns show 
distinct asymmetries which can only be explained be relaxing the requirement of 
symmetric bodies. As a result, the scattering patterns simulated from the 
neural network predictions match the experimentally obtained patterns almost 
perfectly.

\onecolumngrid

\twocolumngrid

\begin{figure}[htb]       
\includegraphics[width=\columnwidth]{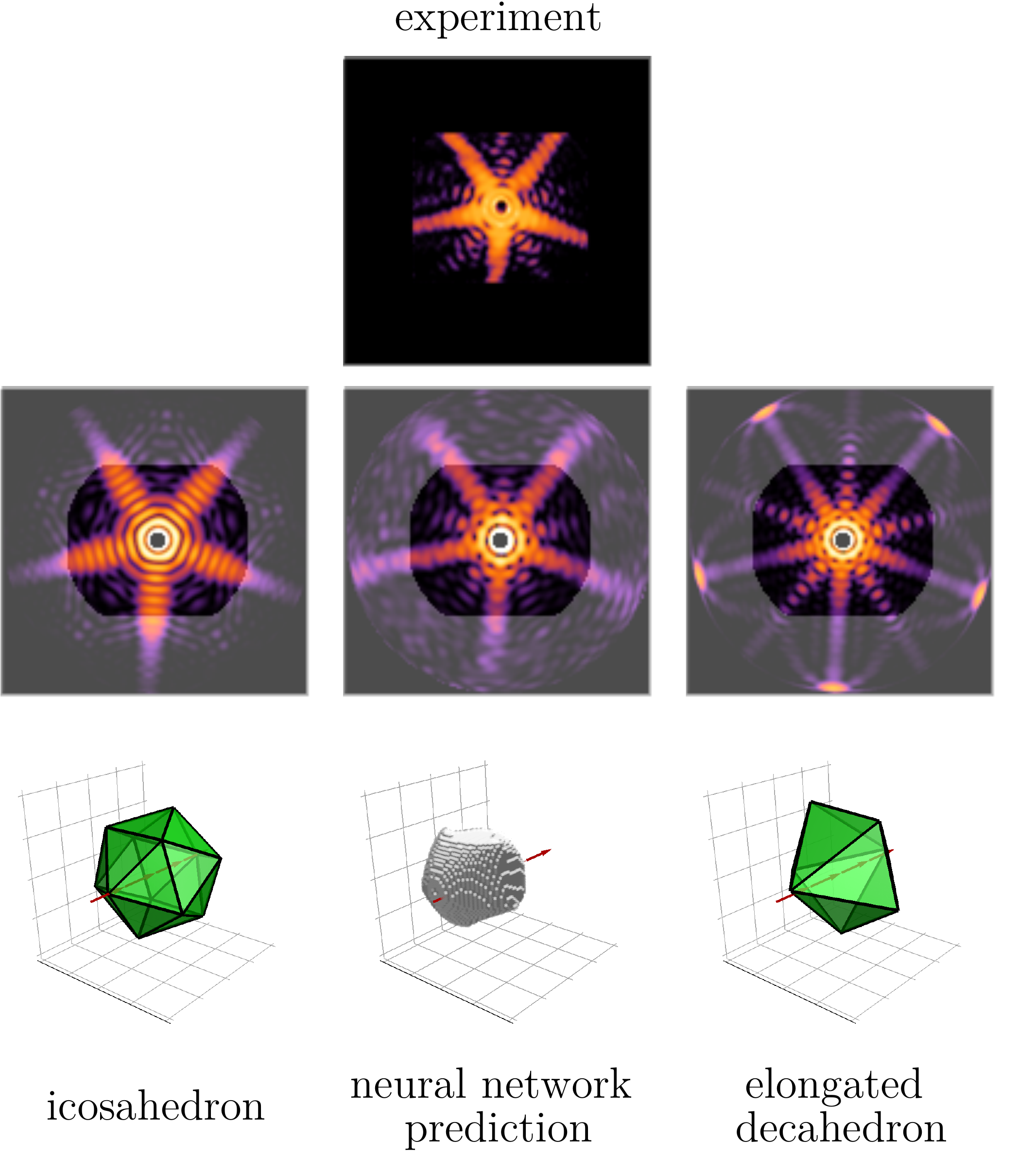}
\caption{The quality of different prediction candidates (bottom row) 
can be judged by comparing the corresponding scattering patterns (center row) 
to the experimental input scattering pattern  (top panel, taken from Ref.~\cite{Barke_2015}, permitted 
by Creative Commons CC-BY 4.0 license (\url{http://creativecommons.org/licenses/by/4.0/})).}
\label{fig:c5Match}
\end{figure}

A particularly striking result is the star-shaped pattern with five-fold 
symmetry (1st column in Fig.~\ref{fig:expPred2}). Previously, this has been 
attributed to an icosahedron (see left panel in Fig.~\ref{fig:c5Match}), 
as this was the only shape in the parametrized 
model with the correct symmetry. 
Instead, the neural network predicts an 
object with a front face resembling an
elongated decahedron of similar size 
(central panel in Fig.~\ref{fig:c5Match}). 
A regular decahedron would produce a 
scattering pattern with ten-fold symmetry. However, the elongation of a
decahedron breaks that symmetry in the wide-angle 
scattering pattern, resulting in two 
distinct sets of five reflexes each with different intensities 
(right panel in Fig.~\ref{fig:c5Match}).  
The reproduction quality of the input scattering pattern can be judged similar to the
scatter loss during training. Within the available detector region, outlined by the grey mask in 
Fig.~\ref{fig:c5Match}, the mean-squared difference between the simulated scattering patterns of the 
prediction candidates and the input scattering pattern is calculated. The reference value is the scatter 
loss of the icosahedron (left panel in Fig.~\ref{fig:c5Match}, forward fitted in Ref.~\cite{Barke_2015})
with a benchmark value of $7.10 \times 10^{-3}$. The physics-informed neural network achieves a much 
closer fit with a mean error value of just $4.63 \times 10^{-3}$.
Starting from the neural network prediction, a decahedron, elongated by a factor of $1.6$ along the 
five-fold symmetry axis, was fitted to the input scattering pattern. In this case, a reproduction
error of $4.74 \times 10^{-3}$ was achieved, which is slightly worse than the neural network candidate
but much closer than the icosahedron.

From the three object candidates shown in Fig.~\ref{fig:c5Match}, the neural network prediction 
achieves the best reproduction of the input pattern. The scattering patterns of all three candidates 
differ strongly outside the detector region. This indicates that the reconstruction quality is mostly 
limited by the available detector range which implies that further progress can only be made be 
enlarging the angular range of the detector.

The physics-informed neural network, however, achieves a much closer fit than the parametrized forward 
fits by successful interpolation between the object classes learned during training. 
The predicted structure, similar to a deformed elongated decahedron, is a novel observation for silver 
nanoclusters of the given size, to the best of our knowledge.
This result 
shows that the neural network reconstruction can help in detecting shapes of 
nanoparticles that would not have been expected from equilibrium cluster 
physics.

\section{Summary}
\label{sec:summary}
We have developed a neural network that is capable of reconstructing 
three-dimensional object densities of silver nanoclusters from single-shot
wide-angle scattering patterns. 
By including the scattering physics into the penalty function,
 overfitting to features not represented within the scattering patterns
 is surpressed. This leads to convergence of the network weights towards a configuration 
 that allows a better interpolation between the object classes of the training set.
It is thus able to predict transient nanocluster structures that would not be 
expected from equilibrium cluster formation theory. Our method is not restricted 
to the example of silver nanoclusters discussed here. The same network 
structure can be used for any system for which the scattering properties (such 
as absorption lengths) are known, and a numerical algorithm to generate 
training data exists. Combined with the fast evaluation times in the $\mu$s 
range, this paves the way to a fully automated reconstruction of the complete 
structure of nanoparticles from single-shot wide-angle scattering images in 
real time.

\section*{Data and code availability}
The source code supporting the findings of this manuscript is
deposited with a  sample portion of each datatset 
within the repository  \cite{code_repo}.
The full dataset is available from the corresponding author upon request.

\begin{acknowledgments}
T. S. acknowledges financial support from ``Evangelisches Studienwerk 
Villigst''. This work was partially funded by the European Social Fund (ESF) 
and the Ministry of Education, Science and Culture of Mecklenburg-Western 
Pomerania (Germany) within the project NEISS (Neural Extraction of 
Information, Structure and Symmetry in Images) under grant no 
ESF/14-BM-A55-0007/19.
\end{acknowledgments}

\appendix*
\section{Comparing supervised and physics informed learning}

The investigation of transient free flying silver nanoclusters with
FEL x-ray sources has led to the discovery of shapes which were not
expected from equilibrium cluster theory. When training a neural network, 
the space of expected structures is determined by the basis of the 
training dataset. The set of basic shapes presented in Fig.~\ref{fig:baseSolids}
consists mostly of highly symmetric objects. This choice was motivated
by the fact that the absorption length of silver with $a_\text{abs} = 12.5 \, \text{nm}$
is much shorter than the considered cluster diameters between 63 and 320~nm. 
Subsequently, the far side of the object has no discernible impact on the scattering pattern.
By training the neural network mostly on symmetric objects, it is expected that the 
predictions of the networks may be completed by symmetric extension from the
regions were direct structure information is available.

\begin{figure}[htb]       
\includegraphics[width=\columnwidth]{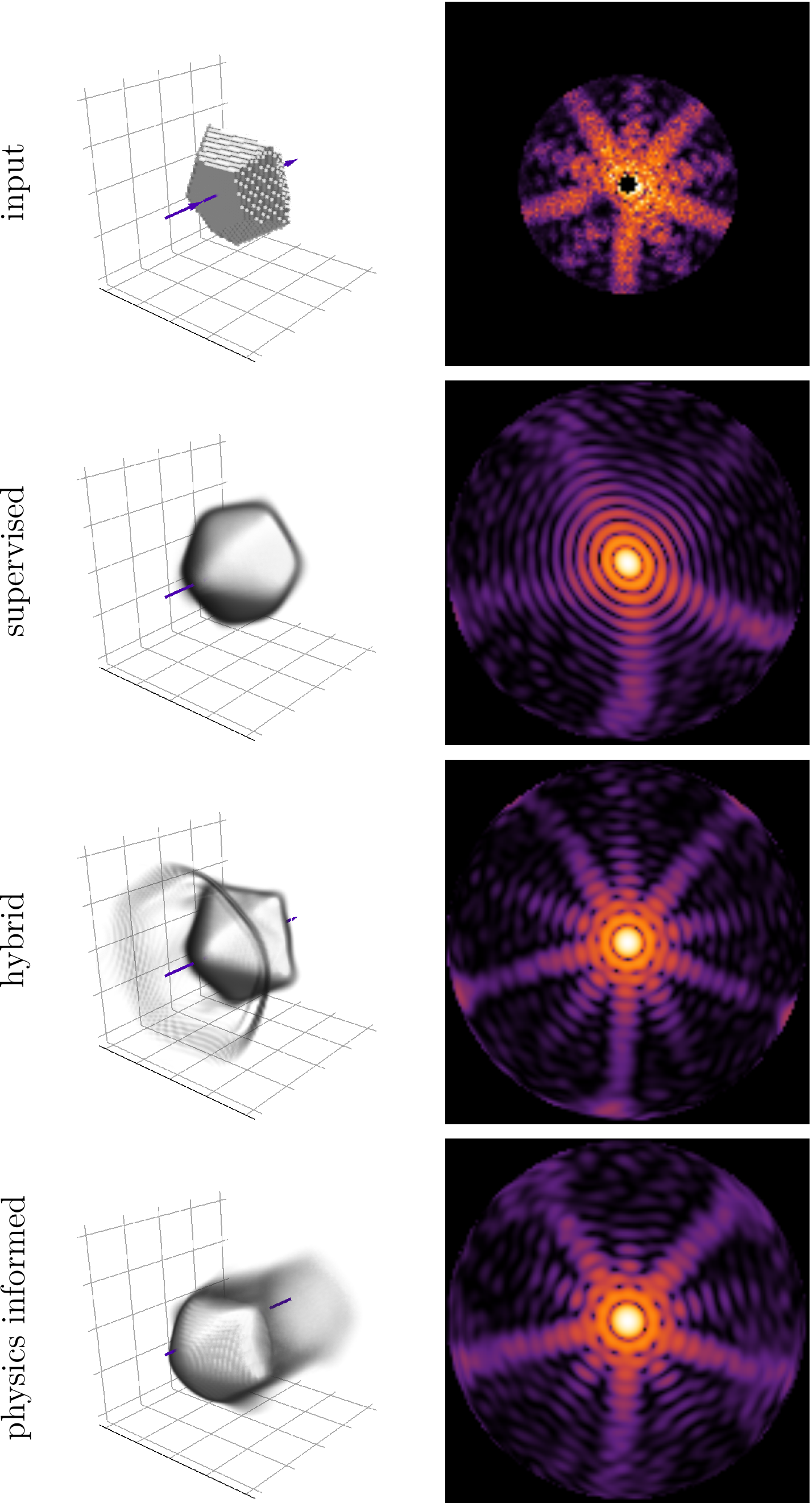}
\caption{The sugmented scattering pattern (top right panel) of an elongated decahedron, cut at half distance along the five-fold symmetry axis (top left panel) is used to test the generalization capabilities of neural networks tested in different training schemes. The left panels show the object predictions made from the test pattern, while the right panels illustrate the corresponding scattering pattterns.}\label{fig:supervisedInformedCompare}
\end{figure}

A key test for the reconstruction network is the generalization capability,
meaning the predictive capacity for objects which are not covered by the 
basis of the training set. For this we created a test set of new objects of deformed base solids by applying
cuts perpendicular to random symmetry axes. An example can be seen in the top row of
Fig.~\ref{fig:supervisedInformedCompare} with an elongated decahedron, cut 
perpendicular to the five-fold symmetry axis at half distance from the center to 
the outer vertex. As long as ideal simulated scattering patterns are used, the neural 
network is capable of reliably reconstructing the objects, independent of the training scheme.
This dramatically changes as soon as image defects are introduced.
An example of this is shown in the second row of 
Fig.~\ref{fig:supervisedInformedCompare}, where the predictions of a neural network
trained by supervised learning are shown. In this approach we used the binary crossentropy
as the loss function within the object space.
The dark areas of the object prediction represent 
regions in which the predicted density takes neither the binary value 0 (no object) 
nor 1 (object), but rather some intermediate value. The reconstructed object differs significantly 
from the ground truth, and the corresponding scattering pattern bears little resemblance to the 
input scattering pattern.

\begin{figure}[bth]       
\includegraphics[width=\columnwidth]{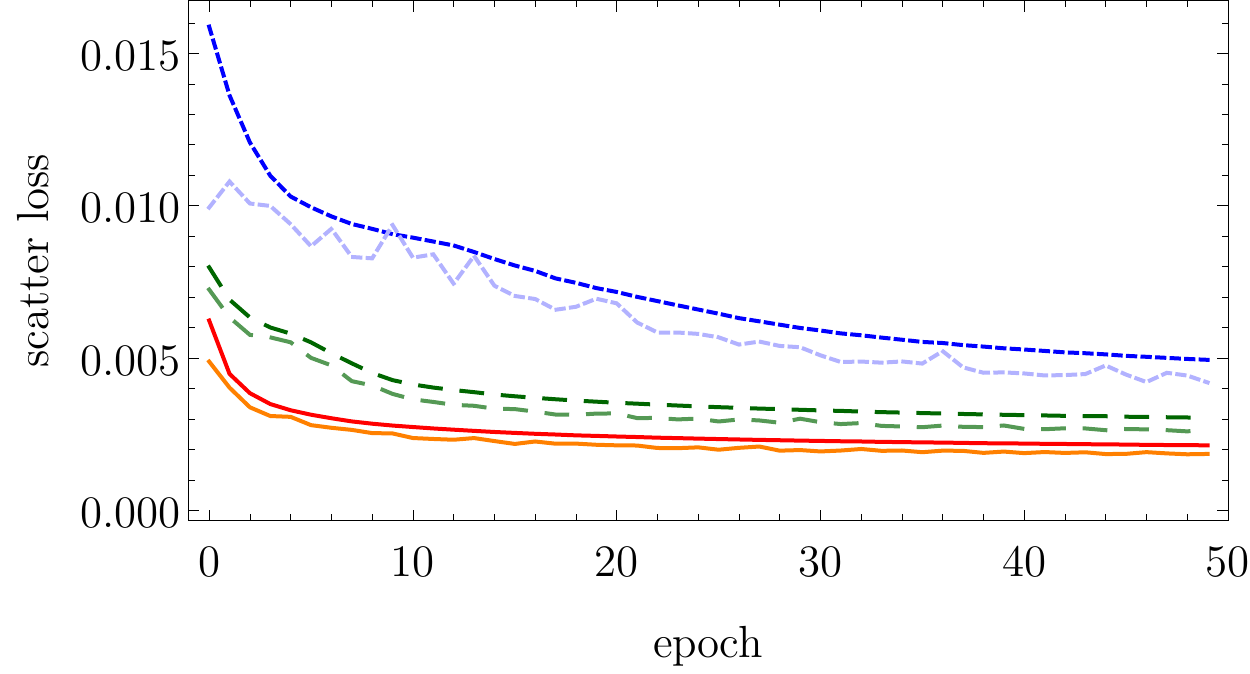}
\caption{Scatter loss recorded over the training of a reconstruction neural network
by binary crossentropy (blue, dotted), hybrid loss (green, dashed) and physics informed training (red, solid). The regular curves
mark the training set performance over each iteration of the training set, while the desaturated curves 
correspond to the validation set.}\label{fig:scatterLossCurve}
\end{figure}

Initially, the scatter loss described in Sec.~\ref{sec:physInf} was developed as
an auxillary loss that is added during supervised training in order to improve the reproduction
quality of the input scattering patterns. 
The result of this modification is depicted in Fig.~\ref{fig:scatterLossCurve} where the scatter loss
recorded during the training of the neural network is shown for both the classical supervised
training (blue dotted curves) as well as hybrid training with auxiliary scatter loss (green dashed curves). 
However, the object predictions are distorted by artifacts, like the ring-structure that can be seen in
the third row of Fig.~\ref{fig:scatterLossCurve}, which are incompatible with the object model.
Balancing the weights between object-space loss and scatter loss in the hybrid loss function 
towards the latter further improves the reproduction quality of the scattering 
pattern during training. This ultimately led us to the implementation of the purely physics-informed 
training (red solid curve in Fig.~\ref{fig:scatterLossCurve}) described in Sec.~\ref{sec:physInf}.

The prediction of a neural network trained purely in a physics-informed manner
is shown in the bottom row in Fig.~\ref{fig:supervisedInformedCompare}. The predicted object
does not correctly reproduce the cut applied to the original object but adds a rounded tip, which
is a known issue, as the associated information within the scattering pattern is easily obfuscated 
by noise (see Sec.~\ref{sec:analysis} for further details). In contrast to the predictions resulting 
from supervised learning, the general structure and footprint of the object are correctly reproduced,
and the corresponding scattering pattern is a good match to the input pattern. These tests were the
decisive criterion in choosing the purely physics-informed training over supervised or hybrid approaches.

\input{discreteNN.bbl}
\end{document}

%% file: discreteNN.bbl
%